\documentclass[fleqn,usenatbib]{mnras}
\usepackage{amsmath}
\usepackage{graphicx}
\usepackage{amssymb,txfonts}

\newcommand{\msun}{{\rm M}_{\sun}}

\newcommand{\xmm}{{\textit{XMM-Newton}}}

\newcommand{\nustar}{{\textit{NuSTAR}}}
\newcommand{\suzaku}{{\textit{Suzaku}}}

\newcommand{\source}{Cyg X-1}

\title[Cyg X-1 in the hard state]
{Analysis of \textit{NuSTAR}\/ and \textit{Suzaku}\/ observations of 
Cyg X-1\\ in the hard state: evidence for a truncated disc geometry}
\author[R. Basak et al.]
{Rupal Basak,$^{1,2}$\thanks{rupal.basak@gmail.com, aaz@camk.edu.pl, mlparker@ast.cam.ac.uk, nislam@camk.edu.pl}, 
Andrzej A. Zdziarski,$^{1\star}$ Michael Parker$^{3\star}$
and Nazma Islam$^{1\star}$\\
$^1$Nicolaus Copernicus Astronomical Center, Polish Academy of Sciences, Bartycka 18, PL-00-716 Warszawa, Poland\\
$^2$Department of Physics, KTH Royal Institute of Technology, AlbaNova University Center, SE-106 91 Stockholm, Sweden\\
$^3$Institute of Astronomy, Madingley Road, Cambridge, CB3 0HA, UK
}
\begin{document}

\date{Accepted 2017 September 01. Received 2017 September 01; in original form 2017 May 18}

\pagerange{\pageref{firstpage}--\pageref{lastpage}} \pubyear{2017}

\maketitle

\label{firstpage}

\begin{abstract}
The geometry of the accretion flow in black-hole X-ray binaries in the hard state, in particular the position of the disc inner edge, has been a subject of intense debate in recent years. We address this issue by performing a spectral study of simultaneous observations of Cyg X-1 in the hard state by {\it{NuSTAR}\/} and {\it{Suzaku}}. The same data were analysed before, and modelled by a lamppost containing hybrid electrons and located very close to the horizon, which emission was incident on a surrounding disc extending almost to the innermost stable circular orbit. We re-analyse the incident continuum model and show that it suffers from the lack of physical self-consistency. Still, the good fit to the data provided by this model indicates that the real continuum has a similar shape. We find it features a strong soft X-ray excess below a few keV, which we model as a soft thermal-Comptonization component, in addition to the main hard thermal-Compton component. This continuum model with reflection of both components yields the overall lowest $\chi^2$ and has a geometry with a hot inner accretion flow and a disc truncated at $\simeq$13--20 gravitational radii. On the other hand, we have also found spectral solution with a lamppost at a large height and a disc that can extend to the innnermost stable circular orbit, though somewhat statistically worse. Overall, we find the fitted truncation radius depends on the assumed continuum and geometry.
\end{abstract}

\section{Introduction}
\label{intro}


Determination of the geometry of inner regions of accreting black-hole (BH) sources is of major importance for our understanding of the physics of those sources. There exist a large number of diverse models of the X-ray emission of those sources, explaining it in terms of accretion discs and their coronae, hot inflows and outflows, and jets. The two main types of those sources are active galactic nuclei and accreting X-ray binaries (XRBs) containing BHs. Here, we concentrate on the latter. 

Luminous BH XRBs show two main spectral states, hard and soft. There is a relative consensus regarding the geometry in the soft state. The main component of the accretion flow is an optically-thick and geometrically-thin accretion disc \citep{ss73,nt73} extending down to the innermost stable circular orbit (ISCO). The hard X-ray tails present in this state with various relative amplitude are probably due to coronal emission containing hybrid, i.e., both thermal and non-thermal, electrons, Compton scattering disc blackbody photons, e.g., \citet{gierlinski99}, \citet{pv09}.

On the other hand, there has been an ongoing dispute regarding the nature of the hard state. The source has to be capable of emitting hard X-rays, which dominate the energetics in this state. A likely radiative process is Compton scattering by a predominantly thermal plasma, given an excellent description of the hard X-ray spectra by it, e.g., \citet{gierlinski97}. A lot of work has been devoted to the model in which this plasma forms a hot accretion flow surrounded by a cold outer disc, e.g., \citet*{SLE_1976}, \citet{ADAF}, \citet*{dgk07}, \citet{yn14}, and many other papers. Such a geometry has to be present in quiescence of BH transients and in initial stages of their outbursts \citep*{lasota96}. In this model, it is natural to associate the transition to the soft state during a transient outburst with the viscous outer disc reaching the ISCO. This model does explain a lot of phenomena observed in BH binaries, e.g., correlations between the X-ray power-law index with the relative strength of Compton reflection, the broadening of the Fe K$\alpha$ line, the characteristic frequencies in the power spectra, and the luminosity, e.g., \citet{dgk07}, \citet{gilfanov10}, \citet*{gzd11}. Also, the main current interpretation for the low-frequency QPOs/breaks in the power spectra is the Lense-Thirring precession of an inner hot flow surrounded by an outer disc (e.g., \citealt*{ingram09,ingram11}), which requires a disc truncation.

However, this model has been questioned in a large number of papers claiming detections of reflection features being as broad as requiring the disc to extend down to the ISCO (or almost to it) also in luminous hard states. Initially, the hard X-ray emission was modelled in those works by a corona above the disc with a prescribed radial emissivity profile of the reflected component, usually steeply rising toward the ISCO (e.g., \citealt*{miller06,rykoff07,tomsick08}). Later, the corona was replaced by a static point-like source on the rotation axis of the BH, so-called lamppost. \citet{mm96}, who introduced this model, noted that there was no reason for all emitting particles to be on the symmetry axis and static, but they used that assumption since it greatly simplified the calculation of the disc irradiation. Fit results using the implementation of this model by \citet{dauser10} often resulted in extreme parameters location of the lamppost, almost on the horizon at the best fit, in \source\ in particular (\citealt{parker15}, hereafter P15).

These results have been controversial. In a number of cases, the same observations as those used by the authors claiming the disc extending to the ISCO have been fitted by others, who found the disc to be truncated \citep*{Done_Diaz_2010,kdd14,plant15,bz16}. The differences have been explained by both instrumental effects (in particular, the pileup) and differences in modelling. In the case of the BH binary GX 339--4, a detailed comparison of different model results is given in \citet{bz16}. In the case of the hard state of Cyg X-1, some works found a truncated disc \citep{done_zycki1999,disalvo01,frontera01,makishima08,yamada13}, and some, a disc extending almost to the ISCO (\citealt*{reis10,miller2012}; P15). 

For the lamppost geometry, \citet*{niedzwiecki16} have pointed out two major problems with its physical self-consistency in the case of the lamppost location close to the BH horizon. Due to light bending, only very few of the emitted photons are able to go to the observer; most either cross the horizon or hit the surrounding disc. This means the actual source luminosity is much larger than that observed. Also, the source is out of equilibrium between production of e$^\pm$ pairs and their annihilation.

Furthermore, the nature of the transition between the hard and soft states becomes difficult to understand if the disc extends to the ISCO in both of them. The disc component in the hard state has a low maximum temperature, $kT_{\rm bb}\sim 0.1$--0.2 keV, and its amplitude is weak, with the Comptonization luminosity being typically a factor of $\sim$10 higher than that in the disc blackbody. This would also mean that the accretion rate through the disc in the hard state is much less than the total one. The state transition would then correspond to the disappearance of the hot plasma and an increase of the disc accretion rate by a factor of a few tens (accounting also for the increase of the luminosity in the soft state) caused by an unspecified factor. No theory of such a process has been proposed yet. Also, the moment when the disc reaches the ISCO in transients, and the hot inner flow/outer disc configuration becomes replaced by a disc-corona geometry (at a luminosity much below that of the state transition) should have some observational signature, while none seems to observed. This lack of an observational signature, in particular no change in the spectral index, is difficult to understand, given the radical change of the energy balance between the hot and cold components predicted in the two cases (e.g., \citealt*{ps96,poutanen97,zls99,pv09}).

Finally, studies of the response of the blackbody component to changes in the hot component flux in the hard state of BH binaries show long time lags, corresponding to the distance between the disc and the bulk of the coronal emission of at least several tens of the gravitational radii \citep{demarco15,demarco17,demarco16}. This is again consistent with the truncated disc geometry.

Given those major issues, we have decided to study again the excellent spectral data from simultaneous \suzaku\/ \citep{suzaku} and \nustar\/ \citep{nustar} observations of \source, investigated before by P15. \nustar, with its unprecedented focusing capability in a wide band covering soft to hard X-rays (3--79\,keV), minimal pile-up, low background and high sensitivity provides a very good quality spectrum which is essential for such studies. In addition, \suzaku\/ extends the energy coverage to both lower and higher energies with a good spectral resolution, which is useful to simultaneously model the disc emission, continuum, reflection hump and the cutoff energy. P15 found the reflecting disc extending down close to the ISCO of a BH with the spin near the maximum value, and irradiated by a lamppost very close to the horizon. They obtained their best fit for the continuum being due to a single source containing hybrid (both thermal and non-thermal) electrons. 

On the other hand, there appears to be strong evidence that the 1--10 keV underlying spectrum is given instead by at least two thermal Comptonization components. First, broad-band spectral fits of \source\ in the hard state of \citet{disalvo01}, \citet{frontera01}, \citet{makishima08} and \citet{nowak11} required a soft Comptonization component in addition to the disc, hard Comptonization and reflection components. Then, \citet{yamada13} have analysed hard-state \source\ data from \suzaku\/ using temporal spectroscopy finding that at least two Comptonization components, hard and soft, are required. These components can be interpreted as an approximation for an inhomogeneous Comptonization cloud, with the spectra hardening with the decreasing radius. This scenario is also compatible with the X-ray hard lags observed in the hard state of \source\/ \citep*{kotov01}, interpreted as propagating fluctuations in the presence of the hardening of locally-emitted spectra. Indeed, \citet*{rgc99} studied Fourier-resolved spectra of \source\ in the hard state and found the spectra harden with the increase of the considered range of Fourier frequency. The same result has been obtained using the timescale-resolved spectroscopy \citep{wu09}. This strongly argues for the X-ray source being inhomogeneous. Therefore, in our analysis we allow for the presence of two separate Comptonization components. 

Our paper is organized as follows. In Section \ref{cyg} we review the parameters of Cyg X-1 and describe the method of analysis. Section \ref{data} describe the data reduction and Section \ref{wind} presents our treatment of the wind absorption. Section \ref{spectral} presents our spectral analysis, where we first compare directly our results with those of P15, and then develop our own approach. We discuss our results in Section \ref{discussion} and present the conclusions in Section \ref{conclusions}.

\section{The parameters of Cyg X-1}
\label{cyg}

\begin{table*}
\caption{The analysed observations of Cyg X-1 with \suzaku\/ and \nustar. The subsets used for spectral fitting are the entire \nustar\/ data, the XIS1 one simultaneous with it (but excluding X-ray dips), and the entire PIN and GSO data. The used value of the Eddington luminosity, $L_{\rm E}$, is for the BH mass of $14.8\,\msun$ and the H fraction of 0.7, the source flux is unabsorbed, and the luminosity, $L$, corresponds to the 0.01--$10^3$ keV flux at the distance of 1.86 kpc assuming isotropy.
}
\begin{tabular}{cclccccc}
\hline
Detector & Start time (UT) & End time (UT) & Effective & Count rate & $F$(4--79\,keV)$^{(a)}$ & $F$(0.01--$10^3$\,keV)$^{(a)}$ & $L/L_{\rm E}$ \\
 &  &  &exposure\,(ks) & s$^{-1}$ & $10^{-8}$\,erg\,cm$^{-2}$\,s$^{-1}$ & $10^{-8}$\,erg\,cm$^{-2}$\,s$^{-1}$ \\
\hline
\hline
XIS1  & 2014-05-20 06:40:34 & 2014-05-22 07:20:40$^{(b)}$ & 12.0$^{(c)}$ & 123.5 & \\
PIN  & 2014-05-19 06:48:19 & 2014-05-22 07:27:16 & 107.0 & 9.7 & \\
GSO  & 2014-05-19 06:48:19 & 2014-05-22 07:27:16 & 107.0 & 11.7 & \\
FPMA & 2014-05-20 05:58:41 & 2014-05-21 08:45:01 & 34.4 & 179.8 &1.83 &9.80 & 0.022\\
FPMB & 2014-05-20 05:58:41 & 2014-05-21 08:45:01 & 35.3 & 163.4 & \\
\hline
\multicolumn{8}{l}{
$^{(a)}$ The unabsorbed flux calculated for the normalization of the FPMA detector and Model 1.}\\
\multicolumn{8}{l}{
$^{(b)}$ 2014-05-21 09:05:15 for the XIS observation simultaneous with the \nustar\/ one.}\\
\multicolumn{8}{l}{
$^{(c)}$ The XIS exposure time simultaneous with \nustar\/ is 6.4 ks, and that after removing the X-ray dips from it is 4.50 ks.}\\
\end{tabular}\\
\label{log}
\end{table*}

Cyg X-1 is one of the first detected X-ray sources \citep{Bowyer1965}. The current distance determinations are $1.86^{+0.12}_{-0.11}$ kpc based on radio parallax \citep{reid2011}) and $1.81\pm 0.09$ kpc \citep{xiang11} based on a study of dust scattering (though the latter discarded three models yielding larger distances as incompatible with the former). It contains a BH with the mass estimated as $14.8\pm 1.0\msun$ \citep{orosz11}. The value of the BH spin appears uncertain, with some measurements giving the dimensionless spin, $a_*$, close to unity \citep{gou2011, fabian2012, tomsick14}, while \citet{kawano17} obtain $a=0.80^{+0.08}_{-0.30}$. The companion, the 09.7 Iab supergiant HDE 226868 \citep{walborn1973} has a relatively uncertain mass of 17--$35\msun$ \citep{orosz11,janusz2014}. The binary inclination is probably in the range of $\sim\! 25\degr$--$35\degr$ \citep{orosz11,janusz2014}. 

The donor emits strong stellar wind, with the mass-loss rate estimated as $\sim 2.5\times 10^{-6}\,\msun$\,y$^{-1}$ \citep{Hutchings1976,gies03}. At the relatively low separation between the stars of $\sim\! 3\times 10^{12}$ cm, X-ray photons emitted close to the BH suffer substantial attenuation in the wind. The column density through the wind is orbitally modulated (e.g., \citealt{wen99,ibragimov05,grinberg15}), and it peaks at the superior conjunction. The observations studied here were done around it, within the narrow range of the orbital phases of the \nustar\/ observation simultaneous with the \suzaku\/ one of 0.90--0.10 (and that for \suzaku\/ HXD of 0.73--0.28). Thus, we need to take particular care of the wind absorption, though we do not expect to see an effect of the orbital modulation. The wind of \source\ consists of a smooth phase and clumps. The clumps passing through the line of sight cause dips and hardenings in the soft X-ray light curves, which frequency peaks at the superior conjunction \citep*{balucinska-church2000,poutanen2008}.

The column density of the interstellar medium (ISM) towards Cyg X-1 has been estimated as $N_{\rm H}\simeq (6\pm 2)\times 10^{21}$ cm$^{-2}$ based on the reddening towards the donor \citep{balucinska95}. The best current estimate of the $N_{\rm H}$ component in the ISM alone appears to be that of \citet{xiang11}, who obtained the values between 4.65 and $4.85\times 10^{21}$ cm$^{-2}$ based on the modelling of the dust scattering toward \source. \citet{tomsick14} then fitted the soft state of Cyg X-1 including absorption by the stellar wind. That state has much stronger soft emission, allowing for a better determination of the column than the hard state. They found $N_{\rm H}\simeq 6.0\pm 0.3$, $6.2\pm 0.2\times 10^{21}$ cm$^{-2}$. Allowing for some additional uncertainty, we constrain our fitted $N_{\rm H}$ to $(4$--$7)\times 10^{21}$ cm$^{-2}$. 

\vbox to 1cm{\vfill}

\section{The data reduction}
\label{data}

Table~\ref{log} summarizes the considered observations. \nustar\/ and \suzaku\/ observed the source 2014 May 20--21 and May 19--22, respectively, though the X-ray Imaging Spectrometer (XIS; \citealt{koyama2007}) of the latter observed it only May 20--22. In the data reduction, we have followed the standard procedures, similarly to P15, but with the calibration and software updated, as specified below.

The \suzaku\/ data have the observation ID 409049010. The available data from the XIS are from the XIS1 unit only, which was the only one operational at that time due to the limited available power. For the spectral analysis, we use only the data simultaneous with the \nustar. The XIS observation was taken in the 1/4-window mode to cope with the high count rate of Cyg X-1. We use the CALDB files of 2016 June 07 and the HEASOFT v.\ 6.19. The task {\tt aepipeline} is used to make the event list, followed by the tasks {\tt aeattcor2} and {\tt xiscoord} to obtain the updated photon positions. The {\tt pileest} task shows a significant pile-up within the inner radius of $1'$ of the point-spread function (PSF), in spite of using the 1/4-window mode. Hence, we use an annular region with an inner and outer radii of $1'$ and $4'$, respectively, for the data extraction. The pile-up fraction is now reduces to only 1.6 per cent when averaged over photons, or 0.7 per cent averaged over pixels. The background is extracted from a rectangular region near the edge of the exposed detector with a similar area as that of the source extraction region. As a part of the source extraction region lies on the inactive region of the detector, we adjust the BACKSCAL keyword of the source spectral files. Finally, the tasks {\tt xisrmfgen} and {\tt xissimarfgen} were used to create the response and ancillary files. We bin the XIS data with an oversampling factor of 3 with the minimum signal-to-noise ratio of 30. There is no available estimate of the systematic error of the XIS data; however, \citet{tsujimoto11} show significant systematic differences in the fitted spectral index up to 9 per cent between the XIS data and those of other satellites operating in the same energy range. Thus, in order to mitigate the effect of the calibration uncertainties on our joint fits to the XIS1 and \nustar/\ data, we opt for a (modest) 0.5 per cent of the systematic error to the XIS1 spectral data, which we add in quadrature. 

The Hard X-ray Detector (HXD; \citealt{hxd}) of \suzaku\/ consists of Si PIN detectors and GSO scintillators; the data from them were extracted using the scripts {\tt hxdpinxbpi} and {\tt hxdgsoxbpi}, respectively. The non-X-ray background for both of the detectors were obtained from the public sites\footnote{\url{ftp://legacy.gsfc.nasa.gov/suzaku/data/background/pinnxb_ver2.2_tuned} and \url{ftp://legacy.gsfc.nasa.gov/suzaku/data/background/gsonxb_ver2.6}}. In addition, the cosmic X-ray background for the PIN was obtained based on the {\it HEAO\/} observations of \citet{gruber1999}. The response files for the PIN and GSO are \verb|ae_hxd_pinxinome11_20110601.rsp| and \verb|ae_hxd_gsoxinom_20100524.rsp|, respectively, and a GSO ancillary file, \verb|ae_hxd_gsoxinom_crab_20100526.arf|, is used. We add a 1 per cent systematic error to the spectra from both detectors \citep{ota08,bautz09}. We use the entire HXD data set (as in P15), given the very modest spectral variability during the observation.

The \nustar\/ data have the observation ID 30001011007. To process them, we used the {\tt NuSTARDAS} v.\ 1.4.1 and the {\tt CALDB} of 2015 March 16. We used {\tt NUPIPELINE} and {\tt NUPRODUCTS} routines to extract the cleaned event files and the spectral products, respectively. The source region was chosen as a circle with a radius $150''$ centred at the position of Cyg X-1, while the background was chosen as a circle with a radius $100''$ away from the source. The data were binned with an oversampling factor of 3 with the minimum signal-to-noise ratio of 50. This binning, different than that for the XIS, was used due to the very different number of counts and resolution between the two instruments. The \nustar\/ spectrum has extremely high signal-to-noise ratio, but relatively poor spectral resolution. The XIS spectrum, on the other hand, has much lower signal but relatively good resolution. It therefore makes sense to apply a higher binning to the \nustar\/ data than the XIS data (as in P15), so that the resolution of Suzaku is preserved and the high signal \nustar\/ spectrum is not too strongly oversampled. There is no available estimate of the systematic error of the \nustar\/ data. However, \citet{madsen15} have shown the existence of substantial calibration residuals in the \nustar\/ response. Also, there are substantial differences in the spectral slope of \nustar\/ and that of other instruments in the overlapping range, see, e.g., fig.\ 6 in \citet{matt14} and fig.\ 4 in \citet{parker16} for those with respect to \xmm. Furthermore, we see a difference in the spectral slope fitted to the \nustar\/ and XIS1 data of $\Delta\Gamma\simeq 0.1$, see Section \ref{two}. Therefore, in order to mitigate the effect of the calibration uncertainties on our joint fits to the XIS1 and \nustar/\ data, we have opted to add a 0.5 per cent systematic error to the \nustar\/ data.

The data and the background of all the detectors are found to be very similar to those obtained by P15. We have used the following energy ranges for the spectral analysis (similar to P15). For XIS, we choose the 1.2--9.0\,keV range but excluding the 1.7--2.5\,keV range, where there are uncertain calibration issues related to the Si K-edge. The HXD PIN data are used in the 20--70\,keV range, which excludes the low energy range in order to avoid thermal noise, but also the 38--43\,keV range, which has known calibration issues, is excluded. The HXD GSO data are used in the 60--300\,keV range. The \nustar\/ energy range is chosen as 4--79\,keV, to mitigate the effect of the X-ray dips, see Section \ref{wind}. We use {\tt XSPEC} v.\ 12.9.0n for the spectral analyses.

\section{The treatment of the stellar wind}
\label{wind}

\begin{figure}\centering
\includegraphics[width=\columnwidth]{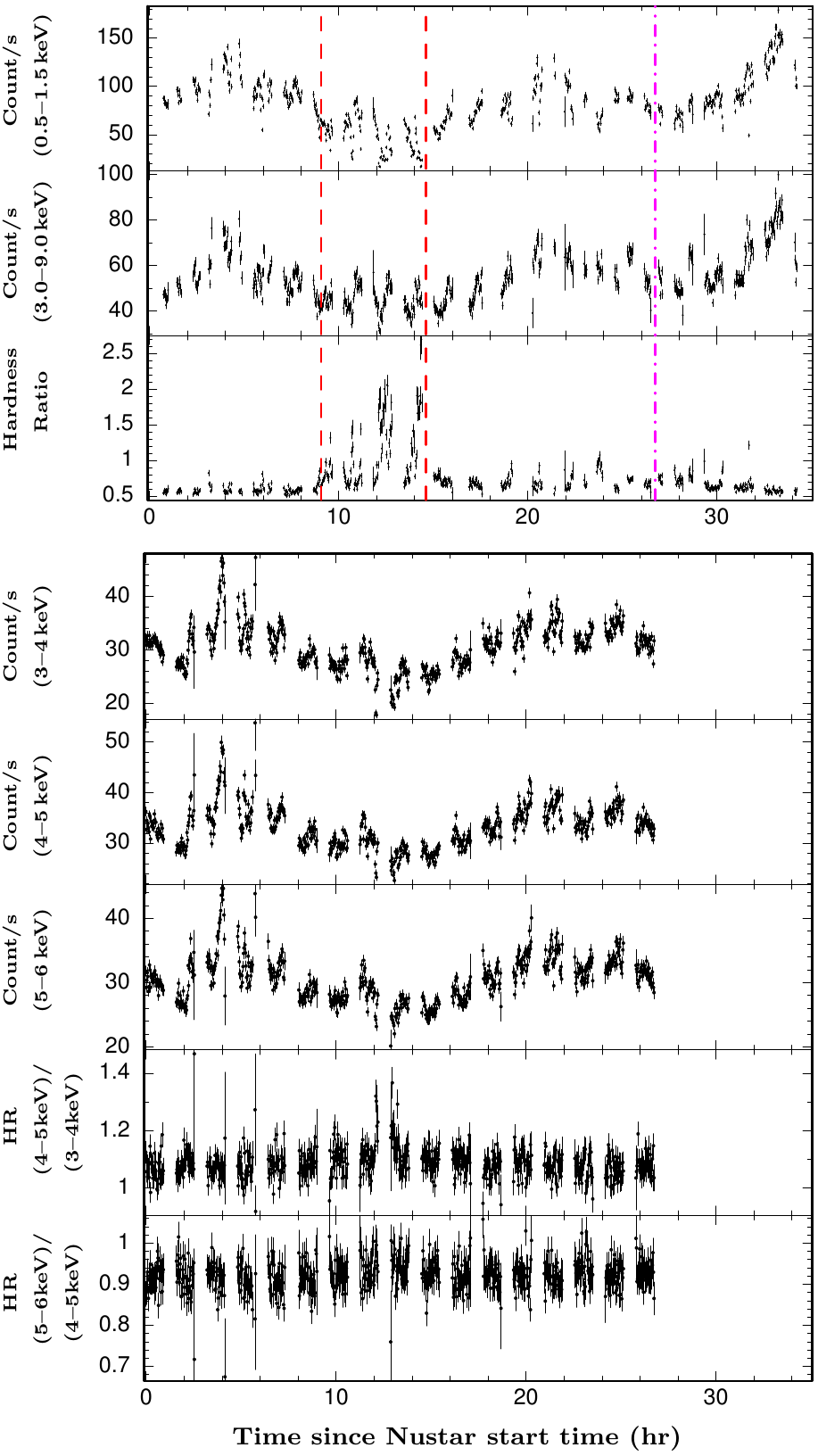} 
\caption{The top three panels show the \suzaku\/ XIS1 light curve in the 0.5--1.5 and 3--9\,keV energy bands and the corresponding hardness ratio. The vertical dashed line show the excluded interval containing X-ray dips, and the vertical dot-dashed line in the \suzaku\/ panels shows the end of the the interval simultaneous with the \nustar\/ observation. The bottom 5 panels show the \nustar\/ light curves in the 3--4, 4--5 and 5--6 keV ranges and the hardness ratios in the 3--5 keV and 4--6 keV ranges. Time is measured from the beginning of the \nustar\/ observation, see Table \ref{log}.
}
\label{lc}
\end{figure}

\begin{figure}
\centering
\includegraphics[width=\columnwidth]{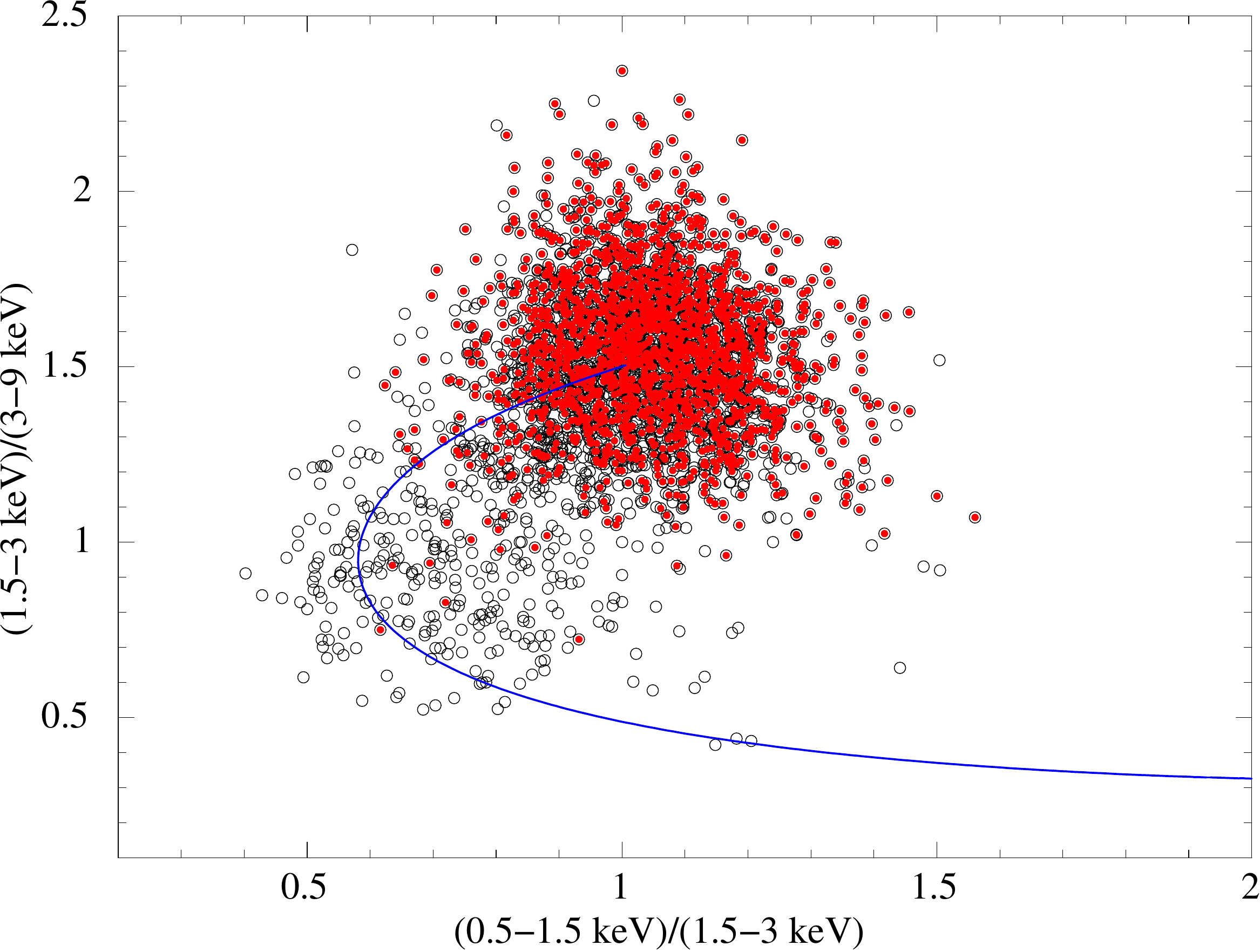}
\caption{The colour-colour diagram for the XIS1 based on the count rates in the 0.5--1.5, 1.5--3 and 3--9 keV energy bands for the observation simultaneous with \nustar\/ (see Fig.\ \ref{lc}) for time bin size of 16 s. The black open circles and red filled ones correspond to the observations during the identified dipping interval (within the dashed lines in Fig.\ \ref{lc}), and outside it, respectively. The latter is used for the spectral fitting. The solid curve shows the positions of points expected using eq.\ (2) of \citet{nowak11} as the H column density of a wind clump changes from 0 (at the top end of the curve) to $2\times 10^{23}$ cm$^{-2}$ (at the bottom end). }
\label{color}
\end{figure}

To take into account absorption by the smooth wind phase, we use a model of an ionized absorber based on the {\tt XSTAR} code version 2.2.1bg (\citealt{xstar}), which was also used by \citet{tomsick14} and P15. This is a table model with the elemental abundances assumed to be solar, the constant H number density of $10^{12}$\,cm$^{-3}$ and the turbulent velocity of $300$\,km\,s$^{-1}$. The free parameters are the ionization parameter, $\xi_{\rm w}$ (in units of erg\,cm\,s$^{-1}$) and the column density, $N_{\rm Hw}$ (in units of cm$^{-2}$), which can be varied in a logarithmic grid in the ranges of $10^2 \leq \xi_{\rm w}\leq 10^5$ and $1.0\times 10^{21} \leq N_{\rm Hw} \leq 5.0\times 10^{22}$. 

We then exclude X-ray dips and spectral hardenings present in the \suzaku\/ soft X-ray light curve due to the clumps. Their frequency peaks around the superior conjunction, which epoch also corresponds to the observations analysed here. The dips can be identified by occurrences of spectral hardenings in the soft light curves. Equivalently, we can use a colour-colour diagram to identify them, following the method of \citet{nowak11}. Those authors noted that \source\ has a significant scattering halo, produced by scattering of its light by dust in the ISM \citep*{overbeck1965, predehl1995, xiang11, smith2016}. This causes an attenuation of the direct flux by scattering of photons away from the line of sight, with the cross section $\propto E^{-3}$. In the case of \source, halo scattering is important at $E\lesssim 3$ keV. For the point-spread function of either XIS or \nustar, this attenuation is compensated by photons emitted to other directions by the source and scattered toward the observer. The light travel time of indirect photons is increased by intervals depending on the scattering position; thus the energy-dependent light curves of halo photons correspond to an average of the past source activity. Therefore, the presence of halo dilutes the X-ray dips. Photons going directly through a wind clump to the observer are attenuated in it, but those coming at the same time from the halo are not. \citet{nowak11} take this effect into account by making simplifying assumptions that the delay-averaged spectrum of \source\ is identical to an instantaneous one and the absorption in the clumps is neutral, see their equation (2). 

In Fig.~\ref{lc}, we show the XIS1 and \nustar\/ light curves and the hardness ratios. We can clearly see dips occurring in the XIS1 data during the time interval of about 9--14 h, when the hardness ratio strongly increases. We exclude this interval from the fitted data, as shown by the dashed vertical lines in Fig.~\ref{lc}. This reduces the effective exposure of the part simultaneous with \nustar\/ to 4.5 ks. Fig.\ \ref{color} shows the colour-colour diagram for these data obtained using the method of \citet{nowak11}. We show the ratio of the XIS1 count rates in the energy bands of 0.5--1.5 keV, 1.5--3 keV and 3--9 keV during the interval with dips and outside it. The solid curve in Fig.\ \ref{color} shows the trajectory expected for dips caused by wind clumps with the H column density between 0 and $2\times 10^{23}$ cm$^{-2}$ obtained using equation (2) of \citet{nowak11}, and assuming the continuum is represented by disc blackbody and a power law absorbed by a neutral medium. We can see that it only approximately determines the part of the diagram affected by the dips, with some dipping events less absorbed in the 0.5--1.5 keV range, which is due to our simplifying assumption of the absorption by a wind clump being neutral. 

Fig.~\ref{lc} also shows that the dipping affects the \nustar\/ light curve only below 4 keV. Given that we have no information about the region below 3 keV, where dipping can be precisely detected, outside the XIS1 observation intervals, we opt here to use the \nustar\/ data above 4 keV only. Summarizing, we use for the spectral fitting the entire \suzaku\/ HXD and \nustar\/ (above 4 keV) data, and the part of the XIS1 data within the \nustar\/ observation, but excluding the identified dipping interval. For those data, we see an approximate constancy of the hardness ratio (Fig.\ \ref{lc}) and thus we do not expect any substantial variability of the absorbing column which could have affected our conclusions regarding the continuum spectral shape.

\section{Spectral analysis}
\label{spectral}

\subsection{Preliminary models}
\label{simple}


The broad-band count-rate spectra and the levels of the background are shown in fig.\ 1 of P15. Since our count-rate spectra look virtually identical, we refer the reader to that figure. Similarly as P15, we first fit the data in a simple way by the Galactic absorption, disc blackbody and thermal Comptonization (using the {\tt nthComp} model; \citealt*{zdziarski96}), with the intention to visualise systematic departures from that model. For the ISM absorption, we use the model {\tt TBnew}\footnote{\url{http://pulsar.sternwarte.uni-erlangen.de/wilms/research/tbabs}} with the abundances of \citet{anders89}. We fix $N_{\rm H}=6\times 10^{21}$ cm$^{-2}$, as in P15. We have found our results to be virtually identical to those shown in fig.\ 1 of P15, with very strong systematic residuals. Therefore, we again refer the reader to that figure. 

Then, we consider the 4--10 keV range from both XIS and \nustar, which we fit with a power-law, absorption and lamppost reflection ({\tt relxilllp} \citealt{Garciaetal_2010, dauser10, Garciaetal_2014}), similarly to P15 (who used the full XIS1 spectrum). They found the disc extending almost to the ISCO, $r_{\rm in}=1.11\pm 0.01 r_{\rm ISCO}$ at $a_*>0.84$, where both the inner and the ISCO radii are in units of the gravitational radius, $R_{\rm g}=G M/c^2$. Here, we use only the XIS1 spectrum simultaneous with \nustar, and obtain results very similar to those of P15. We stress, however, that those narrow-band fits cannot describe the broad-band spectrum.

Indeed, when we consider again the broad-band spectrum, and fit it by thermal Comptonization and Compton reflection, modelled with {\tt relxillCp}, we find very strong systematic residuals, and a large $\chi^2$, with $\Delta\chi^2\simeq +700$ with respect to our best-fit with two Comptonization continua (Model 1 in Section \ref{two}). This clearly rules out a single Comptonization and a disc blackbody as the underlying continuum in \source, in agreement with a number of previous papers, see Section \ref{intro}.

\subsection{The hybrid Comptonization model of Parker et al.\ (2015)}
\label{comparison}

\begin{figure}\centering
\includegraphics[width=\columnwidth]{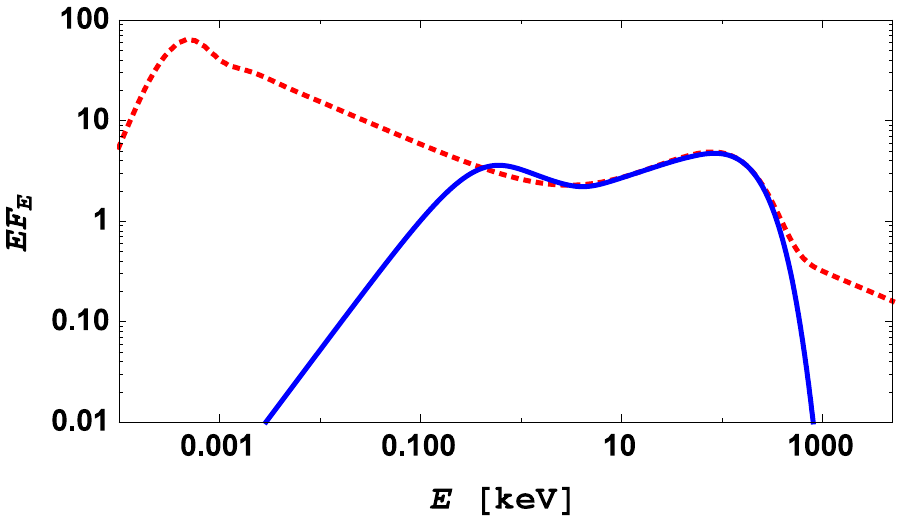}
\caption{The red dotted curve shows the hybrid Comptonization spectrum obtained as the best-fit continuum to the \source\ spectrum by P15. Blackbody photons at $kT=0.11$ eV are Comptonized by a hot plasma, which thermal part has the electron temperature of 70 keV and the Thomson optical depth of 0.57. This yields a power-law spectrum from about 3 eV to 500 eV with the photon index of $\Gamma\simeq 2.4$. The hump beyond that is due to thermal bremsstrahlung, and it is visible because of the very low compactness parameter, equal to $\simeq 0.03$ for the total spectrum (and several times less for photons at $E>1$ keV). Beyond the bremsstrahlung hump, there is a tail due to both bremsstrahlung and single Compton upscattering by non-thermal electrons, at the fitted ratio of the non-thermal to thermal luminosities of $\sim$0.1. The blue solid curve shows our double thermal Comptonization spectrum for Model 1, also shown with its additional components in Fig.\ \ref{models}(b); see Table \ref{t_spectral_fit} for the parameters. 
}
\label{eqpair_nthComp}
\end{figure}

On the other hand, P15 found an excellent fit to the broad-band data using the hybrid Comptonization model {\tt eqpair}\footnote{\url{http://www.astro.yale.edu/coppi/eqpair/eqpap4.ps}} \citep{pc98,coppi99,gierlinski99}. We have refitted that model to the present data and obtained results very similar to those of P15, namely $a_*\simeq 0.98$, $h\simeq 1.1^{+0.8} r_{\rm H}\simeq 1.3^{+1.0}$ (with $1.1 r_{\rm H}$ being the lowest height at which the lamppost model {\tt relxilllp} is defined), $r_{\rm in}\simeq (1.6\pm 0.1) r_{\rm ISCO}\simeq 2.5\pm 0.2$ (which range of radii is somewhat larger than their original result), at $\chi^2\simeq 1232/1000$ d.o.f., where $h$ is the height of the lamppost and $r_{\rm H}=(1+\sqrt{1-a_*^2})$ is the horizon radius\footnote{We note that while P15 stated their values of $h$ are in units of $R_{\rm ISCO}$, they were in units of the horizon radius.} in units of $R_{\rm g}$. The inclination is $i\simeq 48\degr\pm 1\degr$, and the relative Fe abundance is $A_{\rm Fe}\simeq 4.1_{-0.2}^{+0.3}$. 

Unfortunately, there are some major objections to the physical nature of the adopted incident continuum, which spectrum we show in Fig.\ \ref{eqpair_nthComp}. First, the unit of the seed blackbody temperature in {\tt eqpair} is eV, but it was tied to the {\tt diskbb} temperature, which is in keV. Thus, the disc blackbody component had the temperature of 0.11 eV. Second, and more importantly, the total compactness, $\ell\equiv L \sigma_{\rm T}/D m_{\rm e} c^3$, where $L$ and $D$ are the luminosity and source size, was fitted as $\sim\! 2.8\times 10^{-2}$. This compactness implies the X-ray source size as large as $R\sim\! 3\times 10^{11}$ cm. For comparison, $D=0.1 R_{\rm g}$ at the observed luminosity implies $\ell\simeq 5000$, a compactness five orders of magnitude higher than that fitted in P15. Furthermore, most of the model flux was below 1 keV, see Fig.\ \ref{eqpair_nthComp}, while that of \source\ is a small fraction of the bolometric one, which adds one more order of magnitude to the discrepancy. Also, the hard-to-soft compactness ratio was fitted as $\simeq$1.2. These parameters imply that the slope of the thermal Comptonization is $\Gamma\simeq 2.4$, much softer than $\Gamma\simeq 1.6$ found as the overall broad-band slope by P15. This slope is seen at low energies in Fig.\ \ref{eqpair_nthComp}. The hard component in the data is then accounted for by thermal bremsstrahlung, which becomes dominant above $\sim$2 keV, and which is visible only because of the very weak Comptonization cooling at the fitted very low compactness. Obviously, this model cannot correspond to the real physical situation in Cyg X-1\footnote{In fact, as found by \citet{ibragimov05}, no fit with a single hybrid Comptonization model for \source\ can be obtained using realistic constraints on the source size. This was confirmed by P15, who has searched the entire parameter space and found no solution except for the one dominated by bremsstrahlung in hard X-rays. The reason for that is that Comptonization by the thermal electrons always dominates at low energies, while that by non-thermal ones can dominate only at the highest energies, and thus cannot reproduce the observed cutoff at $\sim$100 keV (while it can be responsible for the MeV-range tail, e.g., \citealt{mcconnell02,pv09}).}. Still, its good fit to the overall spectrum reiterates the importance of taking into account the presence of the soft excess in Cyg X-1 (as also discussed in Section \ref{intro}). Fig.\ \ref{models}(a) shows this model with its additional components, disc blackbody and reflection, and Fig. \ref{residuals}(a) shows the fit residuals. We consider below physical models that yield similar underlying continua.

\subsection{Models with stratified Comptonization}
\label{two}

\begin{figure}\centering
\includegraphics[width=0.83\columnwidth,angle=0]{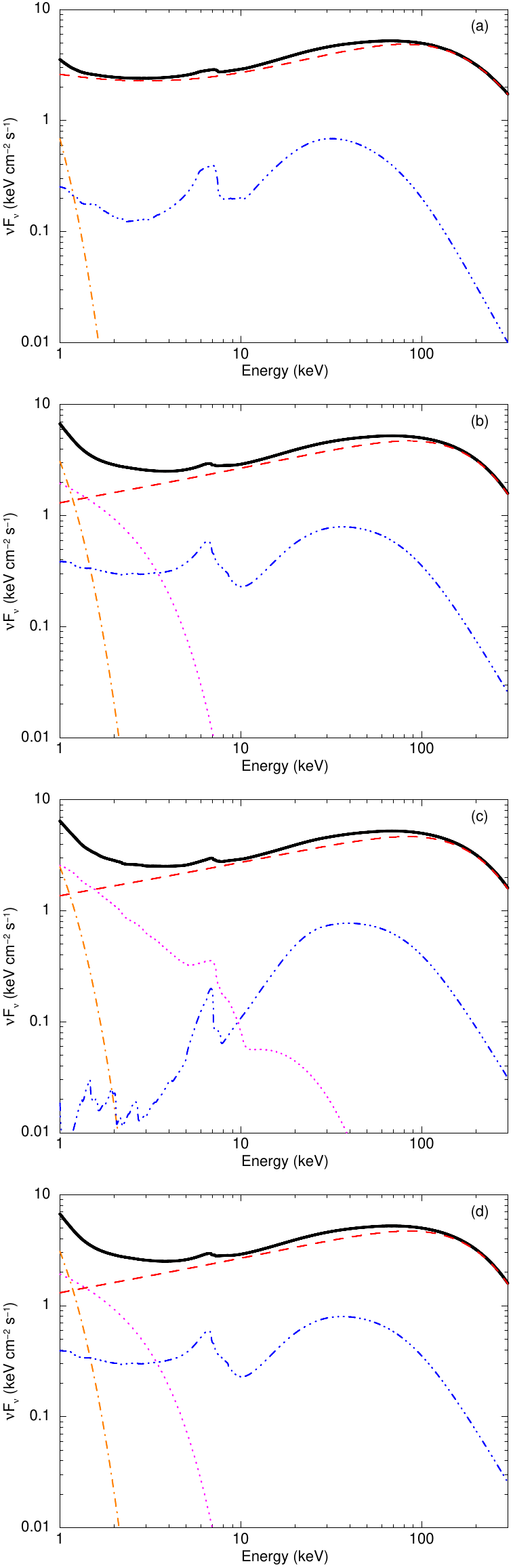} 
\caption{The unabsorbed $EF_E$ spectra of (a) the model of P15; (b) Model 1, (c) Model 2, and (d) Model 4. The (red) dashes show the primary incident continuum, the (magenta) dotted curves show the additional soft continua (the panels b and c only), the (orange) dot-dashed curves show the disc blackbody, and the (blue) triple-dot-dashed curves show the reflection of the primary continua. The (black) solid curves show the sum. 
}
\label{models}
\end{figure}

\begin{figure}\centering
\includegraphics[width=0.84\columnwidth,angle=0]{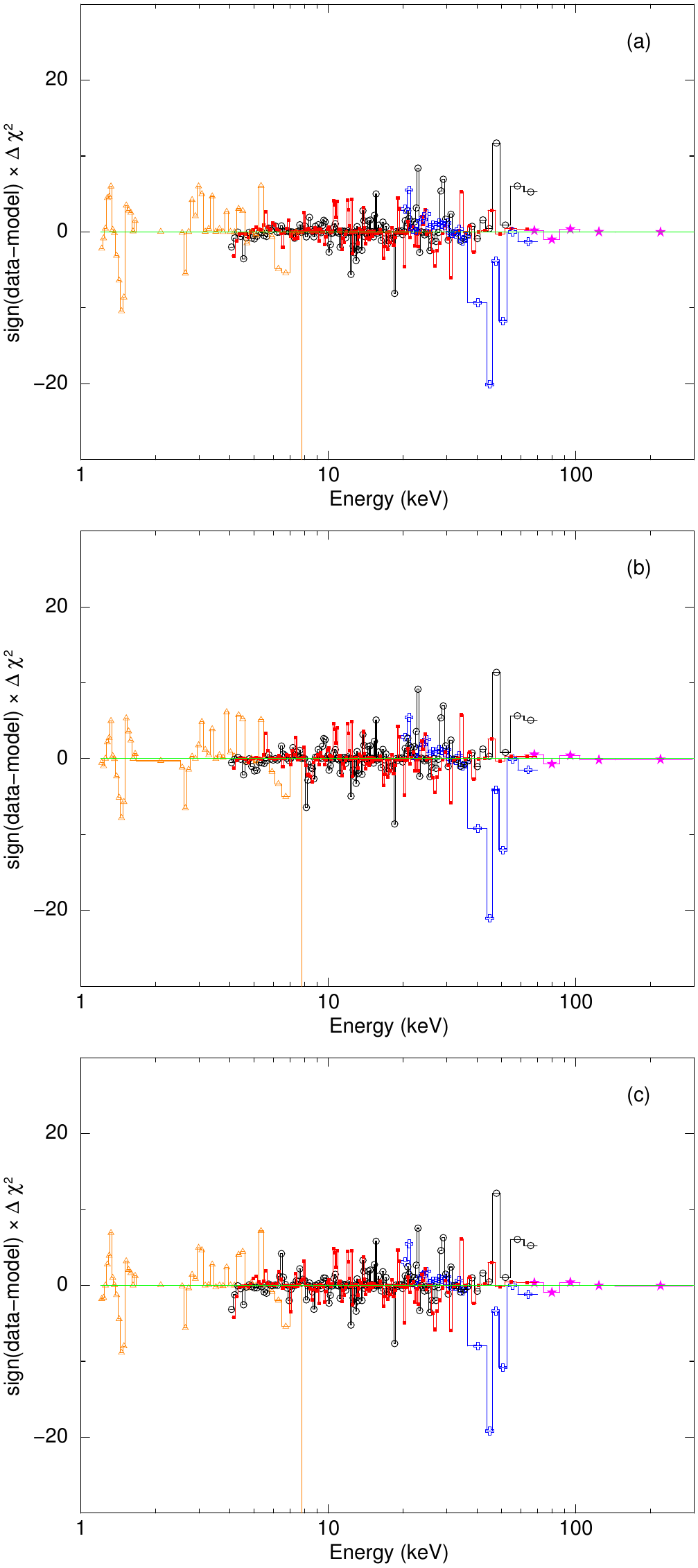} 
\caption{The fit residuals given as $\chi^2$ contributions in (a) the model of P15; (b) Model 1 and (c) Model 2. The colours correspond to different detectors: orange: XIS (1.2--9.0\,keV, neglecting 1.7--2.5\,keV), blue: PIN (20--70\,keV, neglecting 38--43\,keV), magenta: GSO (60--300\,keV), black and red: the \nustar\/ FPMA and FPMB, respectively (4--79\,keV).
}
\label{residuals}
\end{figure}

\begin{table*}
\caption{The results of spectral fitting for our main models. All the models have absorption by the ISM and the wind of {\tt tbnew*xstar}. Their continua are: Model 1: {\tt diskbb+nthComp+relxillCp+nthComp}; Model 2: {\tt diskbb+nthComp+relxillCp+relxillCp}; Model 3: {\tt diskbb+nthComp+relxillCp+relxillCp+xillverCp}; Model 4: {\tt diskbb+nthComp+relxilllpCp+nthComp}.}
\begin{tabular}{cccccc}
\hline
Component & Parameter & Model 1 & Model 2 & Model 3 & Model 4\\
\hline
\hline
ISM absorption & $n_{\rm H}/10^{22}$ & $0.70_{-0.07}^{+0}$ & $0.70_{-0.10}^{+0}$ & $0.70_{-0.08}^{+0}$ & $0.70_{-0.07}^{+0}$\\
\hline
Wind absorption & $n_{\rm Hw}/10^{22}$ & $1.78_{-0.37}^{+0.19}$ & $2.56_{-0.19}^{+0.17}$ & $2.49_{-0.21}^{+0.18}$ & $1.77_{-0.35}^{+0.18}$\\
 & $\log_{10}\xi_{\rm w}$ & $4.34_{-0.05}^{+0.16}$ & $5.00_{-0.08}^{+0}$ & $5.00_{-0.44}^{+0}$ & $4.33_{-0.04}^{+0.16}$\\
\hline
Disc blackbody & $kT_{\rm bb}$ & $0.14_{-0.01}^{+0.04}$ & $0.14_{-0.01}^{+0.02}$ & $0.15\pm 0.01$ & $0.14_{-0.02}^{+0.04}$\\
 & $N/10^6$ & $7.0_{-4.4}^{+2.8}$ & $5.1_{-3.2}^{+7.5}$ & $3.8_{-2.7}^{+4.8}$ & $6.8_{-3.7}^{+24}$ \\
\hline 
Hard Comptonization & $\Gamma$ & $1.70\pm 0.01$ & $1.71\pm 0.01$ & $1.71\pm 0.01$ & $1.70_{-0.002}^{+0.01}$\\
and reflection & $kT_{\rm e}$ & $93_{-6}^{+7}$ & $98\pm 7$ & $98_{-6}^{+9}$ & $94\pm6$\\
 & $N$ & $1.30_{-0.03}^{+0.02}$ & $1.36\pm 0.03$ & $1.36\pm 0.03$ & $1.31_{-0.02}^{+0.01}$\\
& $A_{\rm Fe}$ & $4.4_{-0.3}^{+0.4}$ & $2.4_{-0.4}^{+0.3}$ & $2.2\pm0.2$ & $4.6\pm0.2$\\
 & $\log_{10} \xi$ & $3.28_{-0.07}^{+0.05}$ & $2.00_{-0.16}^{+0.15}$ & $2.00_{-0.13}^{+0.22}$ & $3.28_{-0.07}^{+0.04}$\\
 & ${\cal R}$ & $0.19$ & $0.18$ & $0.16$ & $0.20$\\
& $i$ & $25_{-4}^{+5}$ & $42_{-2}^{+3}$ & $43_{-2}^{+3}$ & $24\pm3$\\
&$r_{\rm in}$ & $6.3_{-1.8}^{+2.2}$ & $17.1_{-2.6}^{+3.5}$ & $15.7_{-2.6}^{+3.2}$ & $1.24_{-0}^{+6.10}$\\
& $h$ &--&--&--& $9.3_{-0.5}^{+1.8}$\\
\hline 
Soft Comptonization & $\Gamma$ & $2.61_{-0.65}^{+0.33}$ & $2.66\pm 0.10$ & $2.65\pm0.08$ & $2.60_{-0.46}^{+0.23}$\\
and reflection & $kT_{\rm e}$ & $0.67_{-0.06}^{+0.15}$ & $4.90_{-0.30}^{+1.10}$ & $5.06_{-0.75}^{+1.01}$ & $0.66_{-0.06}^{+0.07}$\\
& $\log_{10} \xi$ & -- & $4.00_{-0.14}^{+0.18}$ & $4.00_{-0.18}^{+0.10}$ & --\\
 & $N/10^{-2}$ & $195_{-99}^{+48}$ & $2.6^{+0.3}_{-0.6}$ & $2.6\pm0.3$ & $192_{-39}^{+41}$\\
\hline
Static reflection & $\log_{10} \xi$ & -- & -- & $0.68_{-0.68}^{+2.07}$ & --\\
& ${\cal R}$ & -- & -- & $0.02$ & --\\
\hline
Detector normalization & XIS & $0.898\pm 0.004$\\
with respect to FPMA & PIN & $1.277\pm 0.004$\\
 & GSO & $1.163^{+0.013}_{-0.014}$ \\
 & FPMB & $1.018\pm 0.001$ \\
\hline
& $\chi^2$/d.o.f.  & 1249/1004 & 1226/1003 & 1216/1001 & 1246/1003\\
\hline
\end{tabular}\\
{\it Notes:} $n_{\rm H}$ and $n_{\rm Hw}$ are the H column densities (in the unit of cm$^{-2}$) of the ISM and the stellar wind, respectively, with the former constrained to the (5--$7)\times 10^{21}$. $\xi_{\rm w}$ and $\xi$ are the ionization parameters (in the unit of erg\,cm\,s$^{-1}$) of the wind and the reflector, respectively, with the former constrained to $\leq 10^5$. $kT_{\rm bb}$ and $kT_{\rm e}$ are the maximum disc and electron temperature (in keV), respectively, $\Gamma$ is the Comptonization photon power-law index,
$A_{\rm Fe}$ is the reflector Fe abundance in the solar unit, $i$ is the reflector inclination in degrees, and $r_{\rm in}$ and $h$ are the inner disc radius and the lamppost height, respectively, in units of $R_{\rm g}$, and $N$ is a component normalization. Then $\cal R$ is the strength of the hard reflection defined as the ratio of the 20--40 keV flux of the reflected component to the incident one. Models 1 and 4 have no soft reflection, while it is dominant in Models 2 and 3, i.e., the direct flux is negligible. Correspondingly, the normalization is given as either that of {\tt nthComp} or {\tt relxillCp}. The uncertainties are given for 90 per confidence for one parameter of interest, i.e., $\Delta\chi^2=2.71$. The relative detector normalizations and their errors are found to be the same for all of the shown models.The uncertainties marked as $^{+0}$ or $_{-0}$ correspond to the boundary of the allowed region for a parameter.
\label{t_spectral_fit}
\end{table*}

We consider here two thermal Comptonization components, hard and soft, each modelled with {\tt nthComp}. We add relativistically blurred reflection to either only to the dominant hard component (Model 1) or to both (Model 2). We model reflection with either {\tt relxillCp} or {\tt relxilllpCp}, which models use the above thermal-Compton model for the incident continuum. We note that they assume that the maximum temperature of disc blackbody photons (which serve as seeds for Comptonization) is 0.05 keV, without allowing it to vary (J. Garc{\'{\i}}a, private communication). Therefore, we use them for reflection only, and still use {\tt nthComp} for Comptonization spectra. At low energies, we include blackbody emission from the disc, which we model using {\tt diskbb} (\citealt{Mitsudaetal_1984, Makishimaetal1986}). The input seed photons for each of the {\tt nthComp} Comptonization components are assumed to have a disc blackbody distribution with the same temperature as the {\tt diskbb} component. However, the spectrum at energies $<$1 keV is certainly more complex than a simple disc blackbody, in particular due to the strong irradiation of the disc by the hard X-ray source \citep{gierlinski08,gierlinski09}. A study of those effects is outside the scope of this work. Thus, we consider the data at energies above 1.2 keV only, the same as in P15. In {\tt relxillCp} and {\tt relxilllpCp}, we assume the maximum value of the spin parameter, $a_*=0.998$ (cf.\ Section \ref{cyg}), at which value $r_{\rm ISCO}\simeq 1.24$ and $r_{\rm H}\simeq 1.13$. However, given that we find $r_{\rm in}$ substantially above the ISCO in most models, this value has a little effect on our results. The outer radius of the accretion disc is set to $1000\,R_{\rm g}$ and the radial emissivity power-law index is set to 3, though in one model we fit it as well as allow for a broken power-law emissivity profile.  

We first consider the case with no reflection of the soft Compton component (Model 1). The fitting results are shown in Table \ref{t_spectral_fit}. The $\chi^2\simeq 1249/1004$ d.o.f.\ is slightly worse than that of the {\tt eqpair} model, see Section \ref{comparison}. The model yields $r_{\rm in}\simeq 6\pm 2$, which corresponds to a truncated disc at the assumed maximum spin, but it can reach ISCO for lower spin values. We also find $i\simeq 25^{+5}_{-4}\degr$, compatible with the modelling of the binary, see Section \ref{cyg}. The soft-Compton component has $\Gamma\simeq 2.6$ and a very low temperature of $kT_{\rm e}\simeq 0.7$ keV. Fig.\ \ref{models}(b) shows the unabsorbed model components, and Fig.\ \ref{residuals}(b) shows the residuals. In the former figure, we see the strong contribution of the soft-Compton component at low energies, exceeding that of the hard Comptonization below $\simeq$1.5 keV.  

We then add reflection to the soft Compton component (Model 2). This significantly improves the fit, to $\chi^2\simeq 1226/1003$, i.e., substantially better than the {\tt eqpair} model of P15, which has the $\chi^2$ higher by 6 in spite of three more free parameters. The fitting results are shown in Table \ref{t_spectral_fit}. Fig.\ \ref{models}(c) shows the unabsorbed model components, and Fig.\ \ref{residuals}(c) shows the residuals, The model yields $r_{\rm in}\simeq 17\pm 3$, i.e., a significantly truncated disc. The soft-Compton component has $\Gamma\simeq 2.6$ and a low temperature of $kT_{\rm e}\simeq 5.0$ keV. The main differences with respect to the previous model are that now the soft component is reflection-dominated (i.e., the fraction of the direct flux is consistent with null) and both components contribute to the Fe K$\alpha$ line. The dominance of reflection, at face value, requires shielding of a part of the disc. However, it may indicate the presence of a spectral component additional to the two Compton ones of Model 1, and/or some spectral complexity still not reproduced by the current models. The fitted inclination is found to be $i\simeq 42^{+3}_{-2}\degr$. We note that the best-fit model of P15 yields even higher inclination, $i\simeq 48\degr\pm 1\degr$, while the soft-state studies of \citet{tomsick14} and \citet{walton16} yield $i\sim 40\degr$. All these values are significantly larger than the binary inclination of \citet{orosz11} of $27.1\degr\pm 0.8\degr$. On the other hand, the analysis of \citet{janusz2014} points to higher values, $\simeq 30\degr$ or more. Furthermore, it is possible that the inner disc is aligned with the BH spin axis, which in turn may be different from the binary axis. This appears to be the case in the BH binary GRO J1655--40, which has an accurate determination of the orbital axis inclination of $68.7\degr\pm 1.5\degr$ \citep{beer02}, significantly different from the jet inclination of $85\degr\pm 2\degr$ \citep{hjellming95}, with the latter likely to correspond to the BH spin axis. Naturally, systematic inaccuracies of the used model will also contribute to this difference, which, overall, appears to be relatively mild. 

\begin{figure}\centering
\includegraphics[width=\columnwidth]{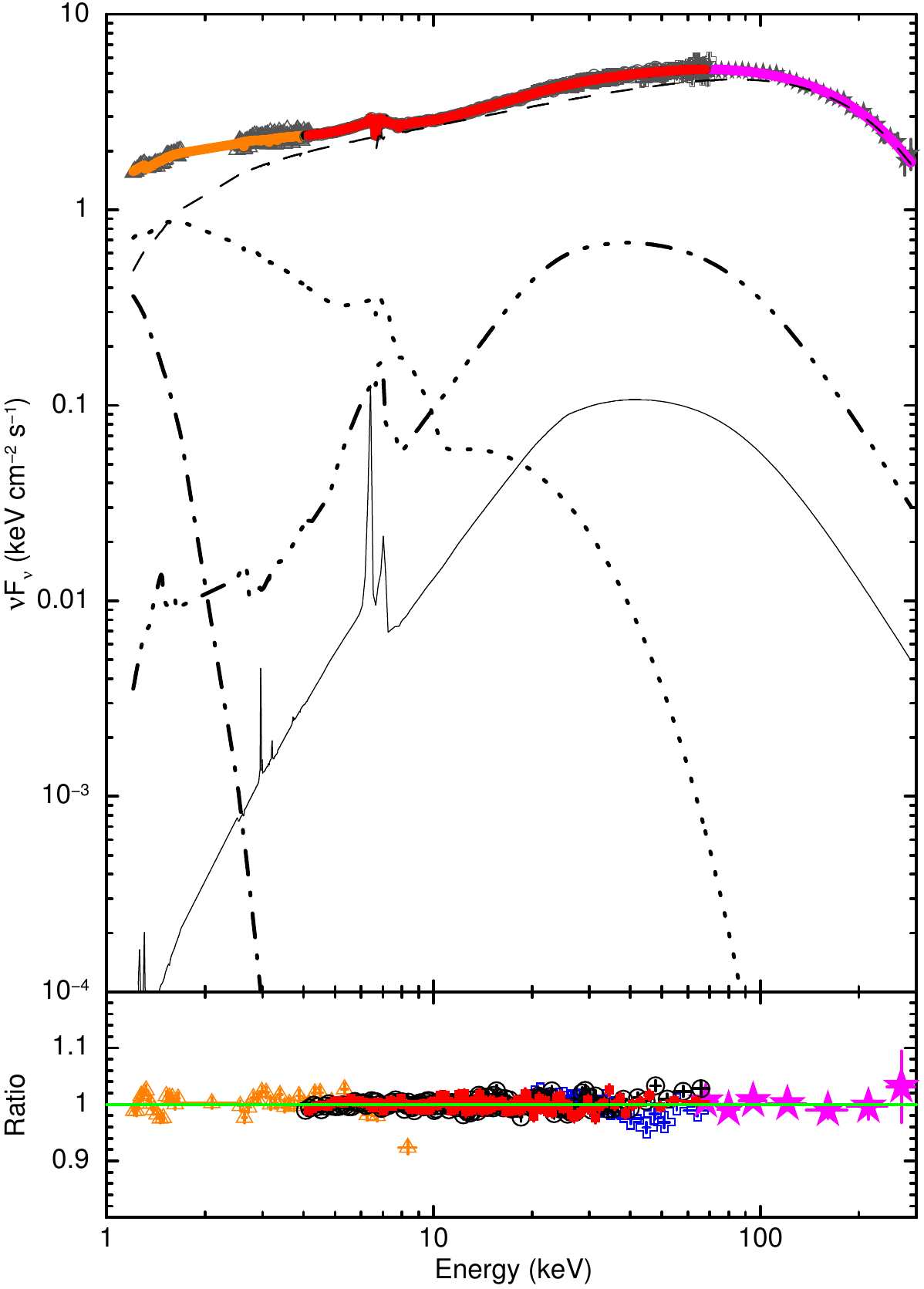} 
\caption{The upper panel shows the unfolded best-fit absorbed $EF_E$ data and model spectrum for our overall best Model 3. The model and data are normalized to those of the FMPA detector of \nustar. Colours in the data indicate the detectors and are the same as in Fig.\ \ref{residuals}. The total model is shown by the thick solid curve, while the dot-dashes, dashes, dots, triple-dot-dashes and thin solid curve correspond to the disc blackbody, the hard Comptonization, the relativistically blurred reflection of the soft Comptonization and of the hard Comptonization, and the static reflection, respectively, all in black. The lower panel shows the fit residuals given as the data-to-model ratio.
}
\label{best_spectrum}
\end{figure}

We have then searched for spectral solutions with a third thermal Comptonization component, which can be an approximation to the case with a continuous distribution of the temperature and the optical depth of the hot plasma. However, we have not found any model with a significant improvement with respect to the double-Compton case. Also, we have studied models with the reflection emissivity index different from the standard value of 3 (which we also adopt here), but we found its best fit is $q\simeq 3.00^{+0.34}_{-0.36}$ for our Model 2. Thus, we have kept $q=3$. We also consider a case with the emissivity profile as a broken power law. We kept the outer index set to 3, and allow free inner index and the break radius. At the best fit, we obtain the break radius of 30, the inner radius of 12.7, and the inner index of $-8.6$, at $\Delta\chi^2=-8$. Thus, the emissivity below the break radius is $\propto r^{8.6}$, which is steeply decreasing towards the BH, and thus shows a weak contribution of the region close to the fitted inner edge to reflection. This result again shows significant disc truncation. 

We then consider the possibility of another reflection component from an outer region of a flared disc. The standard accretion disc \citep{ss73} is flared, with the fractional scale height increasing with the radius, $H/R\propto R^{1/8}$, which effect becomes even stronger in the presence of irradiation. Such an effect is not included in standard reflection codes, such as {\tt relxillCp}, which assume a flat reflector. Since relativistic kinematic effects are negligible for outer disc, we use for it the static reflection model {\tt xillverCp} \citep{Garciaetal_2010}. We add it with $\Gamma$, $kT_{\rm e}$, $A_{\rm Fe}$ and $i$ tied to those of the hard Comptonization and reflection components in Model 2, which we denote Model 3. We show the fitting results in Table \ref{t_spectral_fit} and the unfolded spectrum and the residuals in Fig.~\ref{best_spectrum}. We obtain $\chi^2\simeq 1216/1001$, which improvement of $\Delta\chi^2\simeq -10$ has the chance probability of 0.018 according to the F-test. {\it This our overall best model}. The reflection strength of the static component is an order of magnitude lower than that of the relativistic reflection. We conclude that the addition of the static reflection is statistically required. We see in Table \ref{t_spectral_fit} that its inclusion only slightly changes the other model parameters, in particular $r_{\rm in}\simeq 16\pm 3$.

As our last model, we consider a variant of the Model 1 (the model with a soft Comptonization component without its reflection) with a lamppost, using {\tt relxilllpCp} (Model 4). As in P15, we allow a free normalization of the reflection component. In contrast to the above models, the fit yields a reflecting disc which can either extend down to the ISCO or be relatively far from it, $r_{\rm in}\simeq 1^{+4.9} r_{\rm ISCO}$. However, the height of the lamppost is large $h\simeq 9.3_{-0.5}^{+1.8}$, much above the range obtained by P15. Statistically, the fit is comparable to our Model 1, with $\chi^2\simeq 1246/1003$ d.o.f. The fitted inclination is low, and compatible with that of \citet{orosz11}, but the Fe abundance is high, similar to Model 1. We show the unabsorbed continuum of this model in Fig.\ \ref{models}(d), where we see that the components of this model are indistinguishable from those of Model 1, Fig.\ \ref{models}(b). We discuss the self-consistency of this spectral solution in Section \ref{lamppost}.

The photon power-law index of the hard Comptonization in all models is $\simeq 1.7$, and the electron temperature is $\sim\! 100$\,keV, which are typical values for the hard state of both \source\ and BH binaries in general, see, e.g., \citet{Zdziarskietal_2004a}. We see in Fig.\ \ref{residuals} that there are no significant residuals at high energies, showing no need to consider an addition of nonthermal electrons to fit the present data (though they are clearly needed at higher energies, see, e.g., a comparison of high-energy spectra measured by different instruments in \citealt*{zls12}). The inner disc temperature is found in to be $\sim$0.15\,keV, similar to that found in other studies of \source, e.g., \citet{disalvo01}. However, the normalization of the {\tt diskbb} component is found to be high and only poorly constrained. The reason for the lack of significant constraints is that the energy range used for the analysis is $\geq\! 1.2$\,keV, which does not allow us to constrain it. Furthermore, strong absorption by both the ISM and the stellar wind makes measurements of the unabsorbed soft X-ray spectrum difficult. In all models, the ISM column is fitted at the allowed maximum, $7\times 10^{21}$. A higher value is highly unlikely and hard-state data are not suitable to determine $N_{\rm H}$, see the discussion in Section \ref{cyg}. Allowing a higher column would further increase the strength of both the {\tt diskbb} and soft Comptonization components, and, we note, it could in no case lead to a disappearance of the soft excess. 

The reflection strength, defined as the ratio of the 20--40 keV flux in the reflection continuum to that in the incident continuum, of the hard-Compton component is found to be $\simeq$0.2, typical to the hard state (e.g., \citealt{gierlinski97}). The relative Fe abundance of the reflector is $A_{\rm Fe} \simeq 2.0$--2.7 for Models 2 and 3, which values are are much lower than that obtained for the model of P15, $4.1_{-0.2}^{+0.3}$. We consider the lower value to be more probable. On the other hand, Models 1 and 4 still yield similarly high relative Fe abundance of $\simeq 4.1$--4.8. The ionization of the reflector is much stronger in Models 1 and 4, and thus it significantly contributes to the line broadening, see a discussion in \citet{bz16}. The stellar wind is found to be strongly ionized, $\log_{10}\xi_{\rm w}\sim 5$. Similar values were found with the same model by \citet{tomsick14} and P15.

We note that all our models are fit at the maximum of the allowed ISM column density, $7\times 10^{21}$ cm$^{-2}$. Therefore, we have tested the effect of relaxing that constraint. In the case of models 2 and 3, we have found $N_{\rm H}\simeq 8.6^{+1.3}_{-2.5}$ and $9.3^{+1.4}_{-2.3}\times 10^{21}$ cm$^{-2}$, respectively, i.e., reaching the value of our assumed maximum within 90 per cent confidence, with the $\chi^2$ values becoming lower by only $\sim$3. Also, the values of $r_{\rm in}$ were unchanged. On the other hand, allowing free $N_{\rm H}$ for models 1 and 4 resulted in unrealistically high values of $\simeq 14.5^{+3.1}_{-2.8}$ and $14.9^{+3.1}_{-2.8}\times 10^{21}$ cm$^{-2}$, respectively. Even more unrealistic were the very high amplitudes of the blackbody component and the soft excess. For example, the total flux of the soft Comptonization component became more than twice that of the hard one, i.e., the latter became a hard excess. This confirms the correctness of our constraint on $N_{\rm H}$. The unrealistic fit results at a free $N_{\rm N}$ appear to be related to our modelling of the blackbody component by {\tt diskbb} being too simplified, as discussed above.

We then study the differences between the \suzaku\/ and \nustar\/ data sets. First, we see in Figs.\ \ref{residuals} and \ref{best_spectrum} that the XIS data significantly disagree with those of \nustar\/ above 8 keV (which is also visible in fig.\ 2 of P15). We have fitted the 3--10 keV spectra from the XIS1, FPMA and FPMB by a power law and a Gaussian line model allowing $\Gamma$ to be different for each detector, and while the fitted indices are virtually identical for the FPMA and FPMB of \nustar, the index fitted to the \suzaku/XIS1 data is softer by $\Delta\Gamma\simeq 0.13$. The reason for this is not obvious, but the \nustar\/ spectra are much more reliable in this energy band. This observation was taken towards the end of \suzaku's life, hence only one XIS unit was operational. The spectrum was strongly affected by pileup (which \nustar\/ does not suffer from), meaning the majority of counts have to be discarded. Finally, the calibration of this band is much more reliable for \nustar, where it lies in the middle of the bandpass, than in \suzaku, where it is at the edge of the detector's energy range. This disagreement does not significantly affect the joint fit, as the \nustar\/ spectra dominate the signal in this band. 

We then fit our best Model 3 separately to the data from each satellite. We find indeed significant differences. The most significant ones concern the reflection parameters. For the \suzaku\/ and \nustar\/ data, we find $r_{\rm in}\simeq 2.0^{+0.5}_{-0.8}$ (where the lower limit corresponds to the ISCO), $14_{-3}^{+4}$, and $i\simeq 53\pm 2\degr$, $41\degr$, $A_{\rm Fe}\simeq 4.8\pm 2.2$, 2.3, respectively. As discussed above, we consider the results for \nustar\/ to be much more likely. Also, the inclination and Fe abundance have more likely values for \nustar. Our results also emphasize the presence of systematic differences between the calibrations of different X-ray instruments, which motivates adding systematic errors to the instrumental spectra. It also shows the risk of relying on a single instrument in deriving physical parameters of cosmic X-ray sources. The joint analysis performed above represents our best estimate of the actual parameters of the accretion disc in the hard state of Cyg X-1.

Another obvious difference concerns the detection of the soft Comptonization component. Since our \nustar\/ data are at $\geq$4 keV, that component is only weakly constrained in those data alone. Still, for the \nustar\/ only data, the addition of the soft Compton component results in $\Delta\chi^2=-50$ for addition of four parameters, in the case of Model 3, with a removal of the soft excess resulting in significant positive residuals at $\lesssim$5 keV.

An issue related to the differences in the simultaneous XIS1 and \nustar\/ data is the effect of adding the systematic errors to those data. We have thus performed a fit of our best model (3) without including those errors. We have obtained very similar values of the fit parameters, in particular, $r_{\rm in} = 14.8_{-1.7}^{+1.8}$, $i = 41\degr\pm 1\degr$, $A_{\rm Fe} = 2.38_{-0.34}^{+0.21}$, and $\log_{10}(\xi) = 2.00_{-0.12}^{+0.08}$, at $\chi^2/$d.o.f.\ $=1459/1001$. The value of the ISM column density is now lower, $N_{\rm H}=0.53_{-0.08}^{+0.09}\times 10^{22}$. This difference is related to the difference between the slopes of the XIS1 and \nustar\/ spectra, discussed above. Without systematic errors, the \nustar\/ data, which have a much better statistics but are significantly harder, dominate, which results in a lower value of the absorbing column.

\section{Discussion}
\label{discussion}

\subsection{The soft excess}
\label{soft}

Our work has confirmed the necessity of including a soft excess in models of the X-ray spectra of \source\ in the hard state. We can obtain good fits to the data only with it, and we have checked that the broad-band best-fit model of P15 also includes an underlying continuum with a strong soft excess. Both analyses could not find even a rough fit without it. A likely origin of the soft excess is an inhomogeneity of the Comptonizing medium, as discussed in Section \ref{intro}, with the local spectrum becomes harder with the decreasing distance to the BH. In the present work, we have taken this into account by an addition of a second Comptonization component in our models of time-averaged spectra. While we believe this is the correct overall interpretation, also given the results of previous studies of the soft excess (see Section \ref{intro}), e.g., by \citet{yamada13}, the present best fit has the soft excess dominated by Compton reflection, which is difficult to interpret physically, and which, in our opinion, indicates that it represents only a phenomenological approximation to the real soft excess spectrum.

On the other hand, another contribution to the soft excess is provided by time variability. \citet*{wu10} found that the X-ray spectra of the hard state of Cyg X-1 fluctuate on time scales $\lesssim$1 s, changing their 2--20 keV photon spectral index in the range of at least $\Gamma\simeq 1.6$--1.8 (but likely more), with $\Gamma$ positively correlated with the flux. This variability pattern implies a pivoting, with the pivot energy being at $E_{\rm p}\gtrsim 20$ keV. The average spectrum is then concave, showing a soft excess at low energies, and it is given by \citep{z03},
\begin{equation}
\langle F(E)\rangle \propto E^{1-\langle\Gamma\rangle}\times \begin{cases}
\sinh x/x, &{\rm uniform};\cr
\exp\,(x^2/2), &{\rm Gaussian},\cr
\end{cases}
\label{av_F}
\end{equation} 
where $\langle\Gamma\rangle$ is the average photon index, $x\equiv \Delta_\Gamma \ln(E/E_{\rm p})$, and the distribution of $\Gamma$ is flat from $\langle\Gamma\rangle -\Delta_\Gamma$ to $\langle\Gamma\rangle +\Delta_\Gamma$ in the uniform case, and $\Delta_\Gamma$ is the standard deviation in the Gaussian case. The normalized flux variance for the pivoting variability is \citep{z03},
\begin{equation}
\frac{\sigma^2}{\langle F(E)\rangle^2}=\begin{cases}
x\coth x, &{\rm uniform};\cr
\exp x^2 -1, &{\rm Gaussian}.\cr
\end{cases}
\label{sigma}
\end{equation} 
This qualitatively agrees with the rms in the hard state decreasing in the X-ray range, e.g., \citet{gz05}, \citet{gzd11}. \citet{z03} also note that this type of variability can result in time lags between light curves measured at different energies, which effect has then been calculated by \citet{kf04}.

\begin{figure}\centering
\includegraphics[width=0.9\columnwidth]{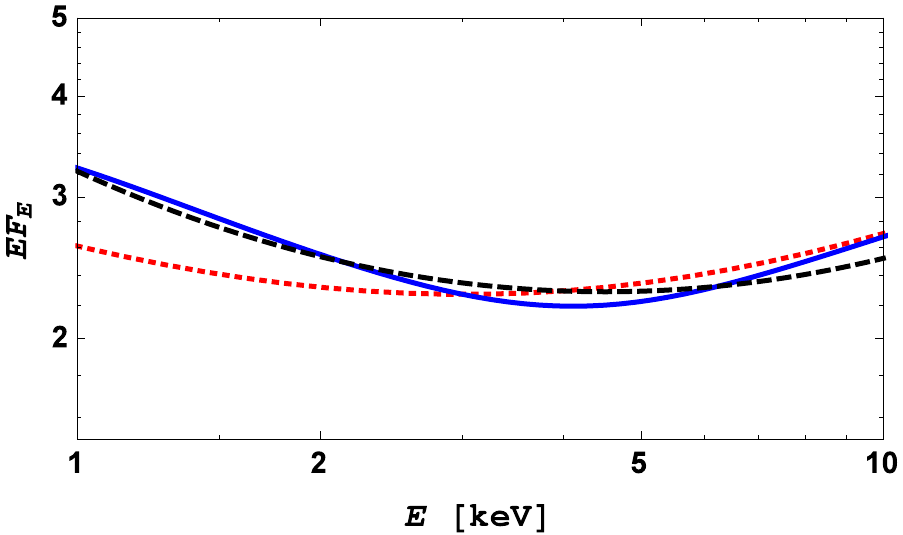}
\caption{Similar to Fig.\ \ref{eqpair_nthComp}, but zoomed to the 1--10 keV range. The red dotted curve shows the hybrid Comptonization spectrum of P15, while the blue solid curve shows the incident spectrum of our Model 1. The black dashed curve shows the variable power-law model of equation (\ref{av_F}), see Section \ref{soft} for the parameters.
}
\label{eqpair_nthComp_zoom}
\end{figure}

In Fig.\ \ref{eqpair_nthComp_zoom}, we show as a red dotted curve an example of the spectrum of equation (\ref{av_F}), for the Gaussian case with $\langle\Gamma\rangle=1.55$, $E_{\rm p}=20$ keV, and $\Delta_\Gamma=0.55$. We see this spectrum is similar to the spectrum of our Model 1. Thus, such variability can in principle reproduce the shape of the X-ray spectrum found in our analysis and modelled by two Comptonization components. In this interpretation, the soft excess is due to the softest power-law components in a pivoting, time-dependent, spectrum, while the hard power law is due to the hardest power-law components. 

In the actual situation, we have both spatial inhomogeneity and fluctuating time dependence, both contributing to the observed curvature in the time-averaged spectra. In fact, $\Delta_\Gamma=0.55$ found here appears larger than $\Delta_\Gamma>0.1$ found by \citet{wu10}. Also, the temperature of the soft excess found here is much lower than that of the hard component, which also argues for the spatial inhomogeneity dominating the soft excess.

\subsection{The lamppost interpretation}
\label{lamppost}

The extreme closeness of the lamppost to the horizon fitted by P15 is unphysical, which general problem is discussed in detail in \citet{niedzwiecki16}. One is a small fraction of photons emitted by a lamppost very close to the horizon reaching the observer, see fig.\ 5 in \citet{dovciak15}. This requires an increase of the accretion rate to compensate for it. The increase is so high that the hard-state accretion rate becomes higher than that in the soft state, which is ruled out for Cyg X-1. In the case of the best fit of this model, with $r_{\rm in}=1.1 r_{\rm H}$, 10 times more photons fall into the BH than reaches the observer. The related problem is that the luminosity measured in the local frame is even higher, due to the photon shift and time dilation effects, and the photon energies in the local frame are strongly blueshifted with respect to the measured energies, $1+z=\sqrt{(h^2+a_*^2)/(h^2+a_*^2-2 h)}$, where $z$ is the redshift measured at infinity. In our best-fit model of P15, we have $1+z\simeq 7.6$. This causes the source to be strongly out of equilibrium between production of e$^\pm$ pairs, produced in photon-photon collisions, and their annihilation. Also, as noted by \citet{niedzwiecki16}, {\tt relxill} and {\tt relxilllp} do not include the redshift of direct photons (only the reflected ones are redshifted), which represents a further problem for the lamppost model of P15. (Fortunately, this does not pose a major problem for our {\tt relxillCp} fits, because the fitted values of the inner disc edge are relatively far from the BH, in which case the redshift is small.) Furthermore, the size of the lamppost in the model of P15 has to be very small, and thus unable to intercept enough disc photons to Comptonize them \citep{dovciak16}.

Another issue is that the theoretical spectrum from the lamppost fitted with the model of P15 has the reflection component more than twice as strong as that observed. For the best fit of this model, the theoretical reflection strength is 0.41, while we observe 0.18--0.20 (Table \ref{t_spectral_fit}). The way the lamppost fit was achieved in P15 was to allow the normalization of the reflected component to be free, and then reduced with respect to the actual lamppost geometry. This is a common practice in lamppost modelling, e.g, \citet{keck15}, \citet{furst15}, but it is clearly not the correct procedure in the case of a specified geometry\footnote{We note that the reflection strength is a free parameter in our fits of coronal models, as well as in other such studies in literature. This is justified by the geometry being not fully specified, e.g., in our case a hot inner flow overlapping with an outer disc, with the unknown amount of overlap and the scale height of the hot flow.}. 

On the other hand, the lamppost may be thought of as an approximation to a vertically stratified source, as discussed in P15. In this case, the observed direct power law can be emitted mostly higher up on the rotation axis, and have a negligible redshift. Still, in order to obtain the observed reflection component from the lamppost part close to the horizon, that part has to emit a strong Comptonization component. Then, the accretion rate has to account for photons swallowed by the BH. This would still result in the accretion rate in the hard state of Cyg X-1 being larger than that in the soft state. Furthermore, an upper lamppost component, which dominates the observed power law in that scenario, would give rise to a strong, weakly-blurred, reflection component, and it is not clear whether a good fit could be obtained with such model. 

The same problem occurs in our lamppost fit, Model 4. The theoretical reflection strength for it is 0.59, while we observe three times less, 0.20 (Table \ref{t_spectral_fit}). Since here the lamppost height is already about $10 R_{\rm g}$, a vertical stratification discussed above would not help. Thus, we consider this model not to be self-consistent.

\subsection{The inner disc radius}
\label{radius}

As discussed above, we have found that although the model of the irradiating spectrum of P15 was valid as a phenomenological continuum, it was not physically self-consistent. That model yielded low values of the truncation radii, $r_{\rm in}\simeq 2.5\pm 0.2$. We have searched for physical models that would yield similar continua. Very similar, but physical, models feature two or more thermal Comptonization components. The models with two components consist of the standard hard thermal Comptonization component, and a soft one, representing the soft excess. Our best Model 3 yields $r_{\rm in}\simeq 13$--20, and it is statistically preferable to that of P15. The specific concave shape of our model is responsible for a reduction of the red tail in the relativistic line profile. This effect has been known for long, probably starting with \citet{wozniak98}, who found it for broad-line radio galaxies.

On the other hand, we have also found that the lamppost model with the hard thermal Comptonization continuum and an additional soft thermal continuum provides a reasonable fit to the data, and yields a low inner radius between the ISCO and $\simeq\! 7 R_{\rm g}$. The height of the lamppost is relatively large, $\simeq\! 10 R_{\rm g}$. However, as discussed above, the reflection strength in that model is three times higher as that observed, arguing against its physical reality.

We note that the relatively large confidence interval of the inner radius in this model is related to the large height of the lamppost. At this height, light bending is weak, and the source is almost isotropic in the observer's frame. Thus, the solid angle subtended by innermost parts of the disc is small. Consequently, the relativistic broadening of the line is moderate, and the line is significantly narrower than in the lamppost model of P15, at which the lamppost was almost on the horizon, and photons were strongly bent toward to the disc. The degeneracy between the lamppost height and the inner radius was pointed out by \citet{fabian14}. In Fig.\ \ref{models}, we also see that there is degeneracy between this model and the coronal model 1. The two look identical in spite of the different inner radii, which is due to the small contribution of the innermost radii to reflection in Model 4, with its large lamppost height.

We note that fig.\ 5 of P15 shows that the Fe K line profiles look very similar in both the hard and soft states. This similarity is interesting, but it can be simply an artefact of choosing a power law as the underlying continuum in the hard state, as done to obtain that figure. That method ignores the presence of a soft excess, which strongly contributes to the claimed red tail of the line, and thus makes is narrower. 

Overall, we interpret our results as showing that the presence of a substantially truncated disc in Cyg X-1 is fully compatible with the data, as well as preferred both statistically and based on physical considerations, see Section \ref{intro}. Still, Fig.\ \ref{residuals}, comparing the residuals of different models, shows that there are no clear systematic differences in the residual patterns. We feel that based on our data alone, we cannot exclude the existence of a physical model yielding a low truncation radius. 

\section{Conclusions}
\label{conclusions}

We have performed a detailed spectral study of simultaneous observations of Cyg X-1 in the hard state by {\it{NuSTAR}\/} and {\it{Suzaku}}. The same data were analysed before, and found to be well modelled by a lamppost located on the axis of a rotating BH very close to the horizon. The surrounding disc was found to extend close to the innermost stable circular orbit. We have found here that the model of the incident radiation used in P15 was not, unfortunately, physically self-consistent, although it was valid as a phenomenological continuum.  

In the present study, we model the incident continuum by two thermal Comptonization components, which reproduce in a plausible way the broad-band spectrum of Cyg X-1 including a prominent soft excess present in this source. Our best model, together with Compton reflection of both components, implies the accretion disc is truncated at $\sim$13--20 gravitational radii. However, we find that the derived inner radius in the hard state is dependent on the continuum model used, and further work is needed to explore physically-motivated multi-component Comptonization models.

We have thus modified some of the conclusions of P15. We find the Fe K line profiles depend on the assumed continuum, and thus their conclusion of the line profile being almost identical in the soft and hard state is model dependent. They state that no models with the disc inner radius $<\!3 R_{\rm g}$ could be found at $3\sigma$ confidence level. We have found a number of such models at a lower $\chi^2$ than that their original best fit. P15 found that only hybrid Comptonization models can fit the \suzaku+\nustar\/ dat. We also find that an addition of hybrid electrons in the Comptonizing plasma is {\it not\/} needed at all in the considered energy range of $\leq\! 300$ keV (while they are needed to reproduce the hard state spectra at higher energies). 

\section*{Acknowledgments}
We thank John Tomsick for providing us with the wind-absorption model used in this work, Thomas Dauser, Chris Done, Javier Garc{\'{\i}}a, Andrzej Nied{\'{z}}wiecki, J{\"o}rn Wilms for valuable discussions, and the referee, Javier Garc{\'{\i}}a, for important suggestions. This research has made use of data obtained through the HEASARC Online Service, provided by the NASA/GSFC, in support of NASA High Energy Astrophysics Programs. RB has been supported by the 2016 START program of the Polish Science Foundation. This research has been supported in part by the Polish National Science Centre grants 2013/10/M/ST9/00729, 2015/18/A/ST9/00746 and 2016/21/P/ST9/04025.

\label{lastpage}

\end{document}